\documentclass[12pt]{article}
\usepackage{amsmath}
\usepackage{times}
\usepackage{graphicx}
\usepackage{color}
\usepackage{multirow}
\usepackage{graphics,epsfig}
\usepackage{amssymb}
\usepackage{epsf,epstopdf,wrapfig}
\usepackage[hidelinks, breaklinks]{hyperref}
\usepackage[authoryear]{natbib}
\usepackage{rotating}
\usepackage{bbm}
\usepackage{latexsym}
\DeclareGraphicsExtensions{.eps,.png}

\newcommand{\captionfonts}{\normalsize}

\makeatletter  
\long\def\@makecaption#1#2{%
  \vskip\abovecaptionskip
  \sbox\@tempboxa{{\captionfonts #1: #2}}%
  \ifdim \wd\@tempboxa >\hsize
    {\captionfonts #1: #2\par}
  \else
    \hbox to\hsize{\hfil\box\@tempboxa\hfil}%
  \fi
  \vskip\belowcaptionskip}
\makeatother

\begin{document}
\hspace{13.9cm}

{\bf \large Predicting single-neuron activity in locally connected networks}

\ \\
{\bf \small Feraz Azhar$^{\displaystyle 1, \displaystyle 2}$}\\
{$^{\displaystyle 1}$Department of Neurosurgery, Brigham and Women's Hospital, Harvard Medical School, Boston, MA, 02115.}\\
{$^{\displaystyle 2}$Department of Neurosurgery, The Johns Hopkins University School of Medicine, Baltimore, MD, 21287.}\\
{\bf \small William S. Anderson}\\
{Department of Neurosurgery, The Johns Hopkins Hospital, Baltimore, MD, 21287.}\\

\thispagestyle{empty}
\markboth{}{NC instructions}
\ \vspace{-0mm}\\
\begin{center} {\bf Abstract} \end{center}
The characterization of coordinated activity in neuronal populations has received renewed interest in the light of advancing experimental techniques which allow recordings from multiple units simultaneously. Across both {\em in vitro} and {\em in vivo} preparations, nearby neurons show coordinated responses when spontaneously active, and when subject to external stimuli. Recent work \citep*{truccolo+al_10} has connected these coordinated responses to behavior, showing that small ensembles of neurons in arm related areas of sensorimotor cortex can reliably predict single-neuron spikes in behaving monkeys and humans. We investigate this phenomenon utilizing an analogous point process model, showing that in the case of a computational model of cortex responding to random background inputs, one is similarly able to predict the future state of a single neuron by considering its own spiking history, together with the spiking histories of randomly sampled ensembles of nearby neurons. This model exhibits realistic cortical architecture and displays bursting episodes in the two distinct connectivity schemes studied. We conjecture that the baseline predictability we find in these instances is characteristic of locally connected networks more broadly considered. 

\section{Introduction}
Quantifying the responses of collections of neurons to the variety of stimuli they are exposed to lies at the heart of deciphering the neural code. As one example, the role of coordinated activity in networks of cortical neurons in particular, represents a natural starting point for the exploration of how neurons encode information relevant to behavior and cognition. Of late, progress has been made towards understanding the responses of small collections of neurons, involved in both lower level sensory processing, and in primary and associative cortical areas, as a result of advances in experimental techniques which make simultaneous recordings from nearby populations of neurons possible \citep*{segev+al_04, gunning+al_05, suner+al_05}. These experimental successes have been coupled with theoretical progress towards understanding collective behavior through the analysis of minimal statistical mechanics based models \citep*{schneidman+al_06, shlens+al_06, tang+al_08}, as well as point process models defined explicitly through maximum likelihood techniques \citep*{brillinger_88, chornoboy+al_88, martignon+al_00, okatan+al_05, nykamp_07, pillow+al_08, truccolo+al_10}. 

In the case of models based on statistical mechanics considerations, joint probability densities of networks of neurons can be constructed assuming maximum entropy distributions consistent with no higher than pairwise correlations (for example), as neurons respond to naturalistic external stimuli \citep{schneidman+al_06}, non-naturalistic external stimuli \citep{schneidman+al_06, shlens+al_06} as well as for neurons undergoing spontaneous activity \citep{schneidman+al_06, tang+al_08}. The resulting models suggest that pairwise interactions capture significant fractions of the information contained in the states adopted by the network. In the case of the point process models, including the intrinsic and extrinsic spiking histories for up to 100ms, for roughly 20--200 neurons, allows one to successfully predict the spiking activities of single neurons in the experimentally measured ensemble \citep{truccolo+al_10}. Notably, in this case, predictability is determined when the external stimuli involves performance of behavioral tasks.  

Coupled together, these results suggest the existence of strongly coordinated activity in small networks of neurons across a broad range of neuronal preparations, in a variety of stimulus paradigms. We investigate whether this coordinated activity can be accounted for by random fluctuations in locally connected model networks. We make progress towards addressing this question using point process modeling in the style of \cite{truccolo+al_10}. We show that in a recently developed computational model of cortex \citep*{anderson+al_09, anderson+al_07}, there exists significant predictive power in small randomly sampled ensembles of neurons driven solely by random background inputs. These inputs drive coordinated activity, which we measure through predictability of single-neuron spiking, and we conjecture that this baseline predictability, which in effect constitutes a lower bound to predictability, is a feature of locally connected networks considered quite generally.

\section{The computational model}

We institute our prediction scheme on a recently developed spiking neural network simulation with realistic cortical architecture \citep{anderson+al_09, anderson+al_07} which displays network bursting dynamics in response to random background inputs. The details of this model have been reported in \cite{anderson+al_09} and \cite{anderson+al_07}, for completeness we collect salient details relevant to our prediction scheme here, as well as in the appendix. 

\subsection{The network}

The version of the model we investigate here has a total of 65,536 single-compartment neurons operating on a modified Hodgkin-Huxley scheme \citep*{avron+al_93}, including spherically symmetric calcium diffusion, and synaptic currents \citep*{kudela+al_03}. The network represents a surface area of 1.6mm $\times$ 1.6mm of simulated cortex, with a thickness of 2.34mm. Neurons are arranged in a repeating lattice of 64 $\times$ 64 minicolumns, with center to center spacing between these minicolumns set at 25$\mu$m. Each minicolumn contains 16 neurons spanning 7 different classes organized in cortical layers. Layer II/III contains 4 excitatory pyramidal cells, 1 inhibitory double bouquet cell, and 1 inhibitory basket cell. Layer IV contains 1 excitatory stellate cell and 1 inhibitory chandelier cell. Layers V and VI each contain 4 excitatory pyramidal cells. Intracolumnar wiring is adapted from visual cortical models and generic somatosensory models \citep{douglas+martin_04, nieuwenhuys_94}. Extracolumnar wiring is based on histological data for the numbers of contacted postsynaptic cells, and is generally isotropic in space within the plane of the postsynaptic cell. The spatial distributions of connections are typically flat out to 200-300$\mu$m, with an exponential tail that can extend to 1mm in the case of inhibitory basket cells.

\begin{figure}
\begin{center}
\includegraphics[width=4in]{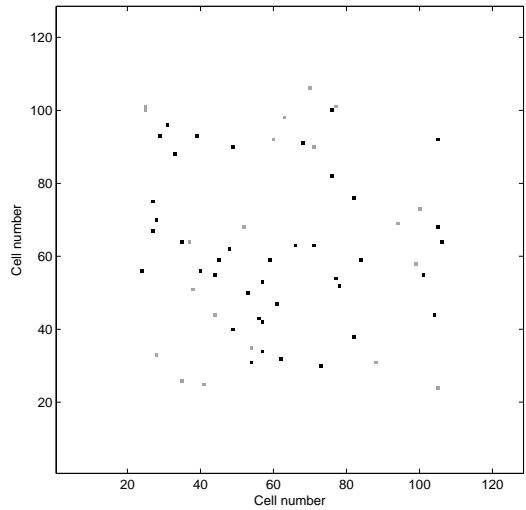}
\end{center}
\caption{Cell locations of the 20 and 40 randomly selected layer II/III pyramidal cell subsets. The 20 cell subset appears in gray, the 40 cell subset in black. The uppermost layer of the model is schematically displayed spanning 128 layer II/III pyramidal cells on each side.}
\label{Fig:Figure_1_CellLocations}
\end{figure}

For the purpose of our prediction scheme, we chose random subsets of 20 and 40 layer II/III pyramidal cells from the central two thirds of the model, representing single-neuron recordings from an area $\sim 1.1\textrm{mm}^{2}$. Figure~\ref{Fig:Figure_1_CellLocations} schematically displays the relative locations of each cell in both of these ensembles.

\subsection{Background input and network activity}\label{subsec:stimulation}

The network is driven by a spatially distributed Poisson background which affects 1\% of the modeled neurons. A synaptic current injection causes these neurons to spike with high probability, with the sequence of input pulses for any one neuron constituting a Poisson process. The baseline synaptic input rate for each affected cell is 2Hz. The background input applied to different neurons is uncorrelated \citep{anderson+al_09}.

The resulting activity alternates in a random fashion between periods of bursting and quiescence. Bursting is characterized by a flat frequency spectrum. Varying the pattern of connections or sequence of background inputs produces similar random activity. Alternations in the frequency of the applied background input, or numbers of connections between cells alters this activity in a graded fashion, with higher connectivity, and higher background input frequency producing constant bursting behavior.

\begin{figure}
\begin{center}
\includegraphics[width=4in]{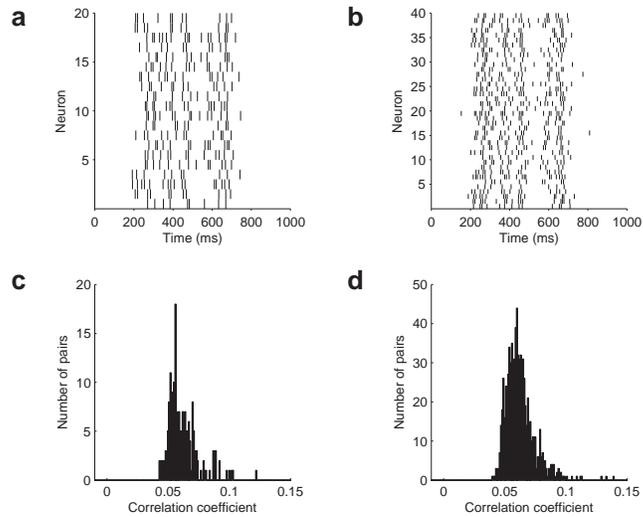}
\end{center}
\caption{Connection pattern 1 spiking behavior: (a) Raster plot showing spike occurrences for 20 randomly selected layer II/III pyramidal cells from the computational model of \cite{anderson+al_09} and \cite{anderson+al_07} during a bursting phase. The burst lasts $\sim 600\textrm{ms}$ and spontaneously terminates. (b) Raster plot showing spike occurrences for the 40 cell subset. (c) Correlation coefficients computed for the 20 cell subset and (d) the 40 cell subset. Following \cite{truccolo+al_10}, they were obtained by considering the extremal Pearson correlation coefficient over all time lags between $\pm 500\textrm{ms}$.}
\label{Fig:Figure_2_TrainsCorr_178}
\end{figure}

In this paper, we probe two separate random patterns of connectivity in the network. In the first, which we will refer to as `connection pattern 1', each layer II/III pyramidal cell contacts at most 178 other layer II/III pyramidal cells out to a radius of 300$\mu$m ($\sim$ 10\% of all layer II/III pyramidal cells encountered within that radius). The resulting network behavior during a bursting period, for both of the 20 and 40 layer II/III pyramidal cell ensembles is displayed in Figure~\ref{Fig:Figure_2_TrainsCorr_178}(a,b) respectively. The model run for connection pattern 1 lasted 330s and was sampled at 1kHz. The extremal Pearson correlation coefficient, defined as that having the greatest absolute value over all time lags between $\pm 500\textrm{ms}$\footnote{\normalsize To be clear, we compute Pearson correlation coefficients for spike counts binned in 1ms bins for each time lag between $\pm 500\textrm{ms}$, and then report the correlation coefficient with the greatest absolute value \citep{welsh_99}.}, is less than 0.15 in either subset.

\begin{figure}
\begin{center}
\includegraphics[width=4in]{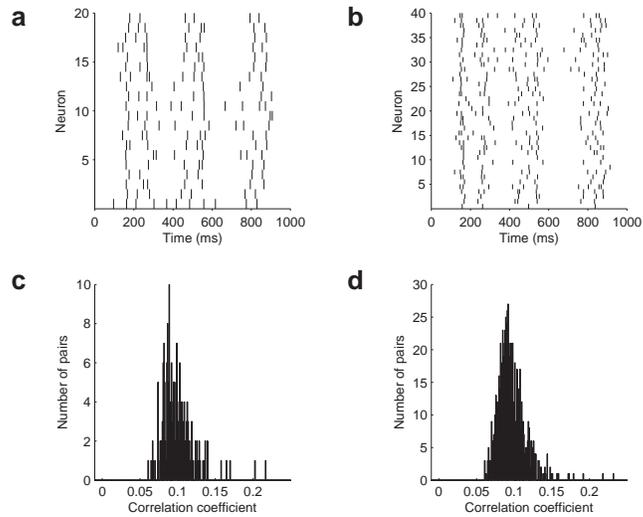}
\end{center}
\caption{Connection pattern 2 spiking behavior: (a) Raster plot showing spike occurrences for 20 randomly selected layer II/III pyramidal cells from the computational model of \cite{anderson+al_09} and \cite{anderson+al_07} during a bursting phase. The burst lasts $\sim 800\textrm{ms}$ and spontaneously terminates. (b) Raster plot showing spike occurrences for the 40 cell subset. (c) Correlation coefficients computed for the 20 cell subset and (d) the 40 cell subset. Following \cite{truccolo+al_10}, they were obtained by considering the extremal Pearson correlation coefficient over all time lags between $\pm 500\textrm{ms}$.}
\label{Fig:Figure_3_TrainsCorr_112}
\end{figure}

In the second pattern of connectivity (`connection pattern 2'), each layer II/III pyramidal cell contacts at most 112 other layer II/III pyramidal cells out to a radius of 300$\mu$m ($\sim$ 6\% of all layer II/III pyramidal cells encountered). Resulting network behavior during a bursting period is displayed in Figure~\ref{Fig:Figure_3_TrainsCorr_112}. The model run for connection pattern 2 lasted 300s and was also sampled at 1kHz. The extremal correlation coefficient in either subset is less than 0.25. We note that intra- and extracolumnar wiring for both patterns of connectivity, for all other pairs of cells was identical.

\section{Inference and prediction formalism}\label{predictionformalism}

The statistical framework within which we propose to perform single-neuron spike prediction is that of point process modeling. In summary, we are interested in writing down a statistical model describing the spiking behavior of a single neuron assuming this model is parametrically related to covariates which we {\em choose}. These parameters are set by optimizing the likelihood of the spike train, namely, by maximizing the joint probability distribution of the spiking activity subject to regularization constraints on the parameters. This joint probability distribution can be written in terms of the {\em conditional intensity function} of the process whose functional form we need to explicitly specify. We outline the pertinent results from the discrete-time point process analysis which we lean on to perform prediction. The reader is referred to \citet*{truccolo+al_05}, \cite{truccolo+al_10}, \citet*{brown+al_04}, \cite{truccolo_chapter}, and \cite{daley+vere-jones_03} for further details.

\subsection{Discrete-time point processes}

Consider an experiment of length $T$ ranging over the time interval $(0,T]$, over which we observe spiking activity for an ensemble of $N_c+1$ cells. We partition the interval into $M$ equal subintervals of length $\Delta$, $\{(t_{i-1},t_{i}]\}_{i = 1}^{M}$, choosing $M$ such that at most one spike falls in any one bin ($M\Delta = T$). We are interested in constructing the joint probability distribution of the observed spike train for a single neuron in our ensemble, neuron $I$ say, and we assume we are free to use information from the other $N_{c}$ neurons in the ensemble to help us. Let $N_{j}^{I}$ represent the number of spikes which have occurred up to and including time $t_{j}$ for neuron $I$, i.e., in the interval $(0, t_{j}]$. Then the quantity
\begin{equation}\label{bincount}
\delta N_{j}^{I}:= N_{j}^{I} - N_{j-1}^{I}
\end{equation}
takes the value 1 if a spike occurred in the time interval $(t_{j-1}, t_{j}]$, or 0 if a spike did not occur in that interval for neuron $I$. Simply put, this counts the number of spikes in the $j\textrm{th}$ bin for neuron $I$. Let $H_{j}$ contain the history of neural spiking for all $N_c+1$ neurons up to but not including the $j\textrm{th}$ bin, and denote the conditional intensity for neuron $I$ in the $j\textrm{th}$ bin by $\lambda_{I}(t_{j}\vert \vec{\theta}, H_{j})$ where ${\vec{\theta}}$ contains the parameters we define the conditional intensity to depend upon. Then one can write the probability of neuron $I$ firing a spike in the $j\textrm{th}$ bin to $O(\Delta)$, as \citep{truccolo+al_05, truccolo+al_10, brown+al_04, truccolo_chapter}, 
\begin{equation}\label{probtoOdelta}
P(\delta N_{j}^{I}=1\vert \vec{\theta}, H_{j})\approx \lambda_{I}(t_{j}\vert \vec{\theta}, H_{j})\Delta. 
\end{equation}
This allows one to write down the joint probability distribution for the spike train for neuron $I$ (i.e., the likelihood) over the observed time interval $(0,T]\equiv (t_{0},t_{M}]$, as
\begin{equation}\label{jointprobs}
P_{I}(\vec{\theta}) = \exp\left(\sum_{j=1}^{M}\left\{\log\left[\frac{\lambda_{I}(t_{j}\vert \vec{\theta}, H_{j})\Delta}{1-\lambda_{I}(t_{j}\vert \vec{\theta}, H_{j})\Delta}\right]\delta N_{j}+\log\left[1-\lambda_{I}(t_{j}\vert \vec{\theta}, H_{j})\Delta\right]\right\}\right).
\end{equation}

\subsection{Logistic conditional intensity function}

What remains to be specified then is the precise functional form of the conditional intensity function $\lambda_{I}(t_{j}\vert \vec{\theta}, H_{j})$, together with the way in which any regularization parameters we may want to use enter into the formalism. 

We follow the spiking histories of the $N_{c}+1$ neurons in $\Delta = 1\textrm{ms}$ bins and consider the influence of previous spikes over a time scale of 100ms \citep[as in][]{truccolo+al_10}. Denoting the spike train of cell $I$ up to but not including the $j\textrm{th}$ bin by $\vec{\sigma}_{j}^{I}$, the histories of our ensemble of $N_{c}+1$ neurons is $H_{j} = \{\vec{\sigma}_{j}^{I}, \{\vec{\sigma}_{j}^{\nu}\}_{\nu = 1}^{N_{c}}\} $. We project the intrinsic spike train (cell $I$) onto $N= 10$ raised cosine functions $\{\vec{B}^{m}\}_{m = 1}^{N}$ of 100ms in length and each of the external cells' spike trains onto $N^{+}=4$ such functions $\{\vec{C}^{n}\}_{n = 1}^{N^{+}}$ \citep{pillow+al_05, pillow+al_08, truccolo+al_10}. We introduce a parameter for each of the raised cosine functions and propose to utilize a logistic form for the conditional intensity \citep{zhao+iyengar_10}. Putting this together, gives
\begin{equation}\label{condint}
\log\left(\frac{\lambda_{I}(t_{j}\vert \vec{\theta}, H_{j})\Delta}{1-\lambda_{I}(t_{j}\vert \vec{\theta}, H_{j})\Delta}\right) = 
\mu + \sum_{m = 1}^{N}\alpha_{m}\vec{B}^{m}\cdot \vec{\sigma}_{j}^{I} +
\sum_{\nu = 1}^{N_{c}}\sum_{n = 1}^{N^{+}}\beta_{n;\nu}\vec{C}^{n}\cdot \vec{\sigma}_{j}^{\nu},
\end{equation}
where the free parameters in the problem $\vec{\theta} = \{\mu, \{\alpha_{m}\}_{m = 1}^{N}, \{\beta_{n;\nu}\}_{n = 1:N^{+};\nu = 1:N_{c}}\}$ number $1+N+N_{c}\cdot N^{+}$, with $\mu$ representing a term related to the background firing rate of cell $I$.

\subsection{Raised cosine functions}

The functions $\{\vec{B}^{m}\}_{m = 1}^{N}$ and $\{\vec{C}^{n}\}_{n = 1}^{N^{+}}$ introduced above, partially overlap over the 100ms of history, with one function in each of the two subsets of functions tapering off to zero just prior to 100ms in the past. They were derived from code associated with \cite{pillow+al_08}, available online at http://pillowlab.cps.utexas.edu/code\_GLM.html. We provide in this subsection, one way to piece together our implementation of the code accompanying \cite{pillow+al_08}, as well as the equations presented therein for generating these functions. 

\begin{figure}
\begin{center}
\includegraphics[width=4in]{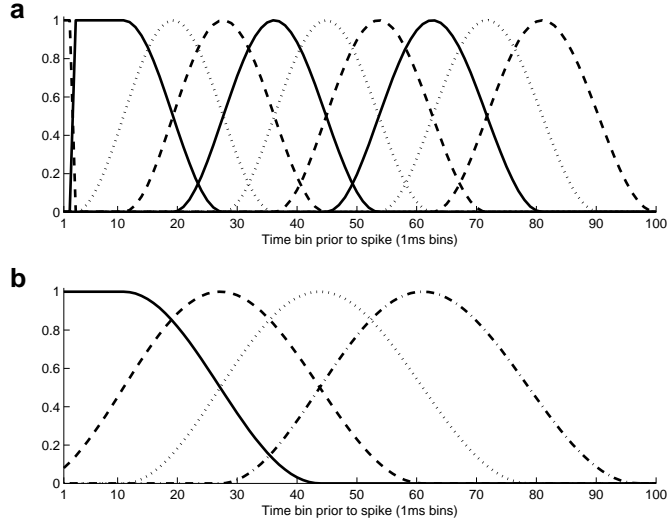}
\end{center}
\caption{Raised cosine functions. (a) For the intrinsic history case, a total of $N = 10$ functions were used onto which the previous 100ms of spiking history was projected for the target cell under consideration. (b) For the extrinsic history case, a total of $N^{+} = 4$ functions were used onto which the previous 100ms of spiking history was projected for each external cell considered in the remainder of the ensemble.}
\label{Fig:Azhar_RCF}
\end{figure}

In the intrinsic history case, we construct a total of $N= 10$ functions. Setting $\epsilon = 0.001$, and discretizing the time axis so that $t= 0, \epsilon, 2\epsilon,\dots,0.1-\epsilon$, gives a total of 100 points which are used to encode the previous 100ms of the neuron's own spiking history (in 1ms bins). A refractory period of 2ms is enforced using the function $B^{1}(t)$ \citep{truccolo+al_10}, where $B^{1}(t)=1\;\textrm{for}\;t = 0, \epsilon$, and $B^{1}(t)=0\;\textrm{otherwise}$. The remaining $N-1$ functions are defined for $j = 2,3 \dots, N $ as $B^{j}(t)= \frac{1}{2}\left[1+\cos\left(a\log(t+c)- \phi_{j}\right)\right]$ for $|a\log(t+c)-\phi_{j}|\leq \pi$, with $B^{j}(t)=0\;\textrm{otherwise}$. Here, $c$ is a parameter controlling the nonlinear stretching of the time axis, $a = \pi (N-2)\left\{2\log\left(\frac{x+c}{y+c}\right)\right\}^{-1}$, $\phi_{j}= a\log(y+c)+\frac{\pi}{2}(j-2)$, where $x$ and $y$ control the positions of the peaks of the last and first raised cosine bumps respectively. Two further amendments are made such that $B^{j}(t)= 0\;\textrm{for}\;t= 0, \epsilon$, again for $j = 2,3, \dots, N $, and $B^{2}(t)= 1\;\textrm{for}\;t= 2\epsilon, \dots, y$. The $N$ resulting 100-element vectors $\{\vec{B}^{m}\}_{m = 1}^{N}$ then correspond to the functions so defined on our discretized lattice, with the choices $c= 0.5$, $y = 0.01$, and $x = 0.08$ (set by hand). Figure~\ref{Fig:Azhar_RCF}a displays intrinsic history functions as generated through the code accompanying \cite{pillow+al_08} (and as directly used in the computations presented in our manuscript).

For the extrinsic history case, we construct a total of $N^{+} =4$ such functions. Again setting $\epsilon = 0.001$, and discretizing the time axis so that $t= 0, \epsilon, 2\epsilon,\dots,0.1-\epsilon$, we form $N^{+}$ functions $C^{n}$, for $n = 1, 2,\dots, N^{+}$, where $C^{n}(t)= \frac{1}{2}\left[1+\cos\left(a'\log(t+c')- \phi'_{n}\right)\right]$ for $|a'\log(t+c')-\phi'_{n}|\leq \pi$, with $C^{n}(t)=0\;\textrm{otherwise}$. Here, $a' = \pi (N^{+}-1)\left\{2\log\left(\frac{x'+c'}{y'+c'}\right)\right\}^{-1}$, $\phi'_{n}= a'\log(y'+c')+\frac{\pi}{2}(n-1)$, for $n = 1, 2,\dots, N^{+}$. We further set $C^{1}(t)= 1\;\textrm{for}\;t \leq y'$. The $N^{+}$ resulting 100-element vectors $\{\vec{C}^{n}\}_{n = 1}^{N^{+}}$ correspond to the functions so defined, with the choices $c' = 0.5$, $y' = 0.01$, and $x'= 0.06$ (again set by hand). Figure~\ref{Fig:Azhar_RCF}b displays corresponding extrinsic history functions as generated through the code accompanying \cite{pillow+al_08}.

\subsection{Optimization scheme}

We set the parameters $\vec{\theta}$ through a scheme effectively implementing maximum a-posteriori estimation with a Gaussian prior over the parameters\footnote{\normalsize We thank Sridevi Sarma for passing along to us a fast truncated L2 regularized binomial logistic regression solver - created by Demba Elimane Ba, MIT Department of EECS, Neuro. Stat. Research Lab (MIT Department of BCS), August 25, 2008.}. In practice, one can think of this as performing optimization over an L2 regularized log-likelihood, with an effective objective function $L_{I}(\vec{\theta}, \eta, \bar\eta)$ given by,
\begin{equation}\label{obj}
L_{I}(\vec{\theta}, \eta, \bar\eta) = \log P_{I}(\vec{\theta}) - \eta\sum_{m = 2}^N\alpha_{m}^2 - \bar\eta \sum_{\nu = 1}^{N_{c}}\sum_{n = 1}^{N^{+}}\beta_{n;\nu}^2, 
\end{equation}
where $\eta$ represents a parameter controlling the regularization of the parameters related to cell $I$'s intrinsic history, and $\bar\eta$ controls the regularization of the parameters related to the $N_{c}$ extrinsic histories. We note that we do not regularize over parameter $\alpha_{1}$ - this constant premultiplies $\vec{B}^{1}$ which is chosen to enforce the 2ms refractory period in the model \citep{truccolo+al_10}. For the sake of computational efficiency, we set the parameters $\eta$ and $\bar\eta$ by hand, effectively probing a subspace of the support of the objective function Eq.~(\ref{obj}). In particular, we chose $\bar\eta >> \eta$ (with $\bar\eta= 10^3\eta$) to probe the regime where the external cells in the model served as perturbations to our underlying predictive capability based solely on a neuron's own intrinsic spiking history. In this way our results represent lower bounds on predictive capabilities in this theoretical setting. Once we have chosen the optimal parameters in this way, one can use Eq.~(\ref{probtoOdelta}) to compute the predicted probabilities of spiking. 

\section{Establishing a lower bound on predictability}

\begin{figure}
\begin{center}
\includegraphics[width=4in]{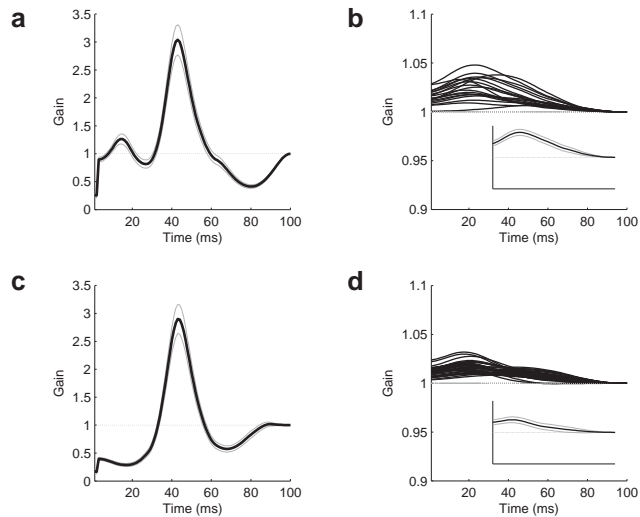}
\end{center}
\caption{Connection pattern 1 exponentiated temporal filters. (a) Intrinsic history filter for a single cell (cell 12) in the 20 cell subset. The $x$-axis denotes time prior to the time bin of interest and the $y$-axis reflects the change in the baseline probability of firing, were a spike to have occurred at $x$ ms in the past. A gain of greater than 1 denotes an increase in the probability of spiking beyond the baseline probability of spiking, a gain of less than one denotes a decrease in this spiking propensity, and a gain of 1 denotes no change in this spiking propensity. An effective refractory period is evident, together with an increase in spiking probability if a spike occurs in the range of $\sim$ 35-60 ms, as well as a longer term depression of spiking probability based on intrinsic activity from $\sim$ 60-100 ms prior to the time bin of interest. (b) Extrinsic history filters for the remaining 19 cells from the 20 cell subset. The greatest increase in the spiking activity of the target cell (cell 12) is controlled by spikes in the ensemble occurring at $\sim$ 20-30 ms prior to the bin of interest. (c,d) display the analogous filters for cell 31 from the 40 cell subset. In each subplot, solid gray lines indicate estimates for uncertainties derived from the standard error associated with parameter fitting. The insets in (b,d) display uncertainties for a single chosen extrinsic temporal filter.}
\label{Fig:Figure_4_Filters_178}
\end{figure}

\begin{figure}
\begin{center}
\includegraphics[width=4in]{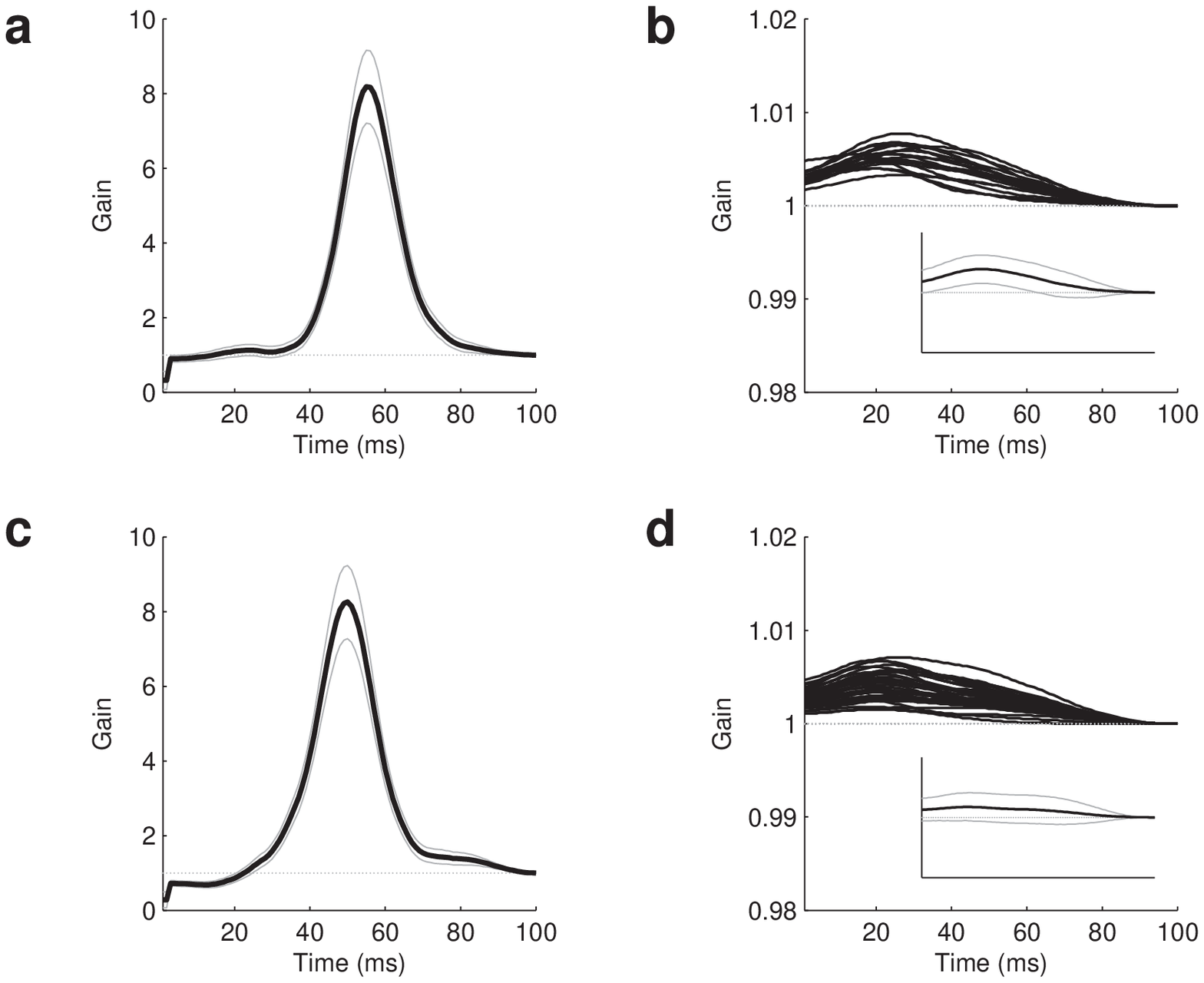}
\end{center}
\caption{Connection pattern 2 exponentiated temporal filters. (a) Intrinsic history filter for a single cell (cell 12) in the 20 cell subset. The $x$-axis denotes time prior to the time bin of interest and the $y$-axis reflects the change in the baseline probability of firing, were a spike to have occurred at $x$ ms in the past. A gain of greater than 1 denotes an increase in the probability of spiking beyond the baseline probability of spiking, a gain of less than one denotes a decrease in this spiking propensity, and a gain of 1 denotes no change in this spiking propensity. An effective refractory period is evident, together with an increase in spiking probability if a spike occurs in the range of $\sim$ 35-80 ms. (b) Extrinsic history filters for the remaining 19 cells from the 20 cell subset. The greatest increase in the spiking activity of the target cell (cell 12) is controlled by spikes in the ensemble occurring at $\sim$ 20-40 ms prior to the bin of interest. (c,d) display the analogous filters for cell 31 from the 40 cell subset. Solid gray lines represent uncertainty estimates as discussed in the caption to Figure~\ref{Fig:Figure_4_Filters_178}.}
\label{Fig:Figure_5_Filters_112}
\end{figure}

The prediction scheme described in section \ref{predictionformalism} leads to estimates for the parameters of the problem $\vec{\theta} = \{\mu, \{\alpha_{m}\}_{m = 1}^{N}, \{\beta_{n;\nu}\}_{n = 1:N^{+};\nu = 1:N_{c}}\}$. Once these parameters have been set, one can compute the intrinsic and extrinsic history filters which quantify the influence of prior spikes on the probability of spiking for any target neuron. Figure~\ref{Fig:Figure_4_Filters_178} displays schematic temporal filters for a single neuron out of the 20 cell subset, and a single neuron out of the 40 cell subset for connection pattern 1. The parameters which determine these displayed filters are as computed for one fold of the $k$-fold cross validation schemes we used to generate results in this paper ($k = 11$ for connection pattern 1, $k = 10$ for connection pattern 2). We display the analogous plot for connection pattern 2 in Figure~\ref{Fig:Figure_5_Filters_112}.

\begin{figure}
\begin{center}
\includegraphics[width=4in]{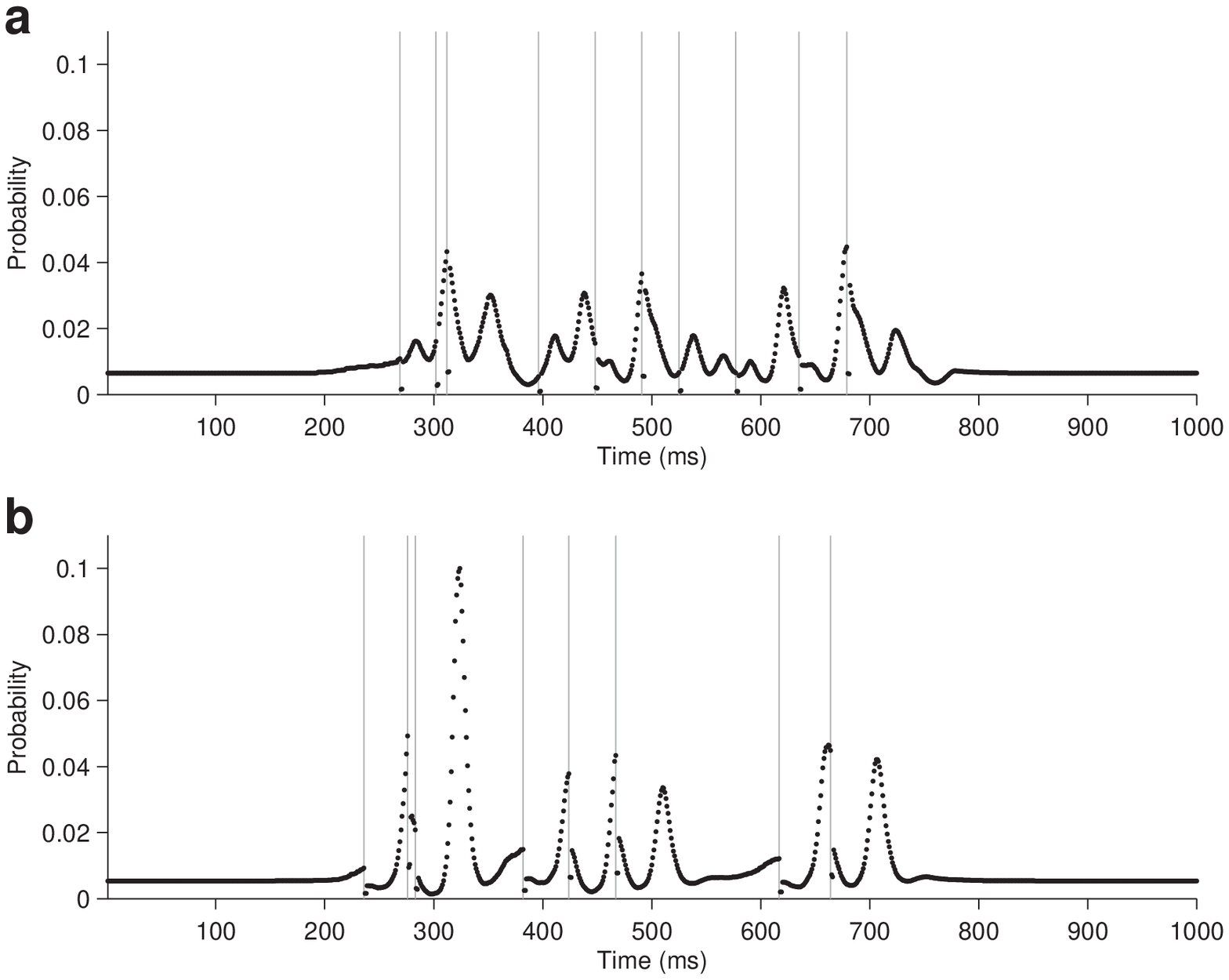}
\end{center}
\caption{Connection pattern 1 conditional probabilities: (a) Predicted conditional probabilities of spiking for target cell 12 from the 20 cell subset, displayed over the same period of time as in Figure~\ref{Fig:Figure_2_TrainsCorr_178}(a). The occurrence of a spike in the target cell is indicated by a vertical gray line. The baseline firing rate is evident in the case where no spikes have occurred in the past 100ms. One example of the combined intrinsic and ensemble activity indicating the increased likelihood of an upcoming spike in the target neuron, is evident in the $\sim$ 20ms prior to the last spike shown on the plot. (b) Predicted conditional probabilities of spiking for target cell 31 from the 40 cell subset displayed over the same period of time as in Figure~\ref{Fig:Figure_2_TrainsCorr_178} (b).}
\label{Fig:Figure_6_condprobs_178}
\end{figure}

\begin{figure}
\begin{center}
\includegraphics[width=4in]{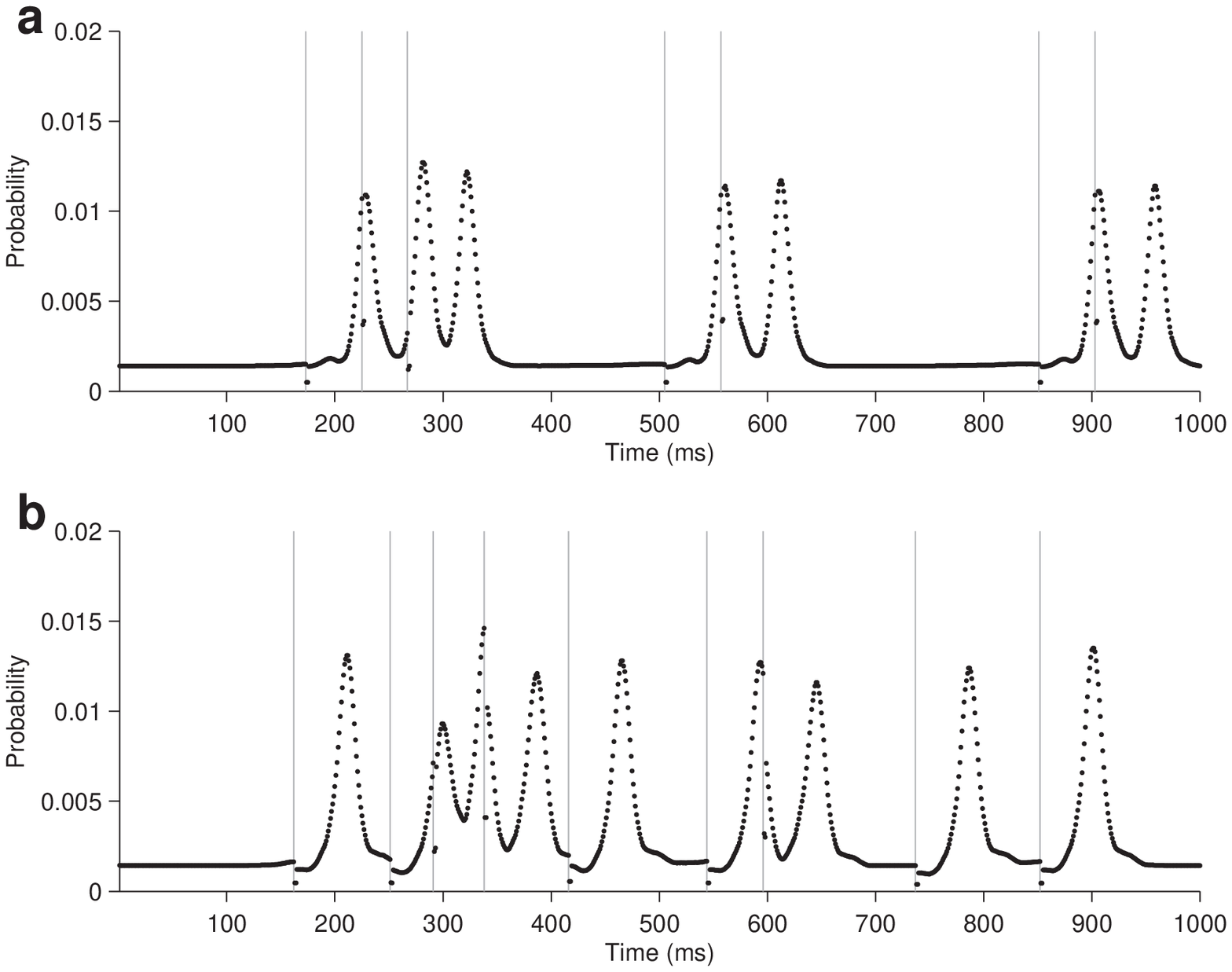}
\end{center}
\caption{Connection pattern 2 conditional probabilities: (a) Predicted conditional probabilities of spiking for target cell 12 from the 20 cell subset, displayed over the same period of time as in Figure~\ref{Fig:Figure_3_TrainsCorr_112}(a). The occurrence of a spike in the target cell is indicated by a vertical gray line. The baseline firing rate is evident in the case where no spikes have occurred in the past 100ms. (b) Predicted conditional probabilities of spiking for target cell 31 from the 40 cell subset displayed over the same period of time as in Figure~\ref{Fig:Figure_3_TrainsCorr_112}(b).}
\label{Fig:Figure_7_condprobs_112}
\end{figure}

Applying temporal filters to our assumed form for the conditional intensity function Eq.~(\ref{condint}), and substituting the resulting expression into that for the probability of spiking Eq.~(\ref{probtoOdelta}), gives predicted conditional probabilities for spiking given recordings over the previous 100ms of activity in the ensemble. An example of the conditional probabilities thus generated, for both connection patterns is presented in Figures~\ref{Fig:Figure_6_condprobs_178} and \ref{Fig:Figure_7_condprobs_112}.

\begin{figure}
\begin{center}
\includegraphics[width=4in]{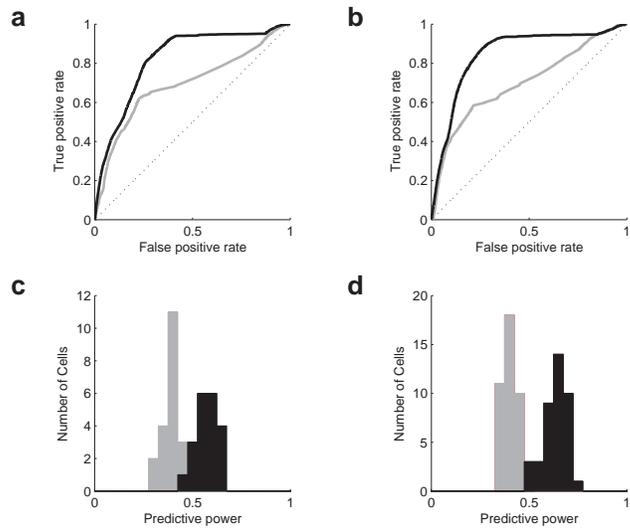}
\end{center}
\caption{Connection pattern 1: (a) ROC curves for our prediction scheme for a single cell (cell 12) from the 20 cell subset. (b) ROC curves for a single cell (cell 31) from the 40 cell subset. The gray trace in both cases results from thresholding over conditional probabilities generated by considering only the intrinsic spiking dynamics of the target cell. The black trace shows the case where the external cells are included in the analysis. (c,d) Histograms of the predictive powers obtained for the 20 cell subset (c), and the 40 cell subset (d). The gray bars indicate prediction based solely on the intrinsic spiking history of the target cell, whereas the black bars indicate predictive powers obtained by including the remainder of the ensemble.}
\label{Fig:Figure_8_ROC_PPHist}
\end{figure}

\begin{figure}
\begin{center}
\includegraphics[width=4in]{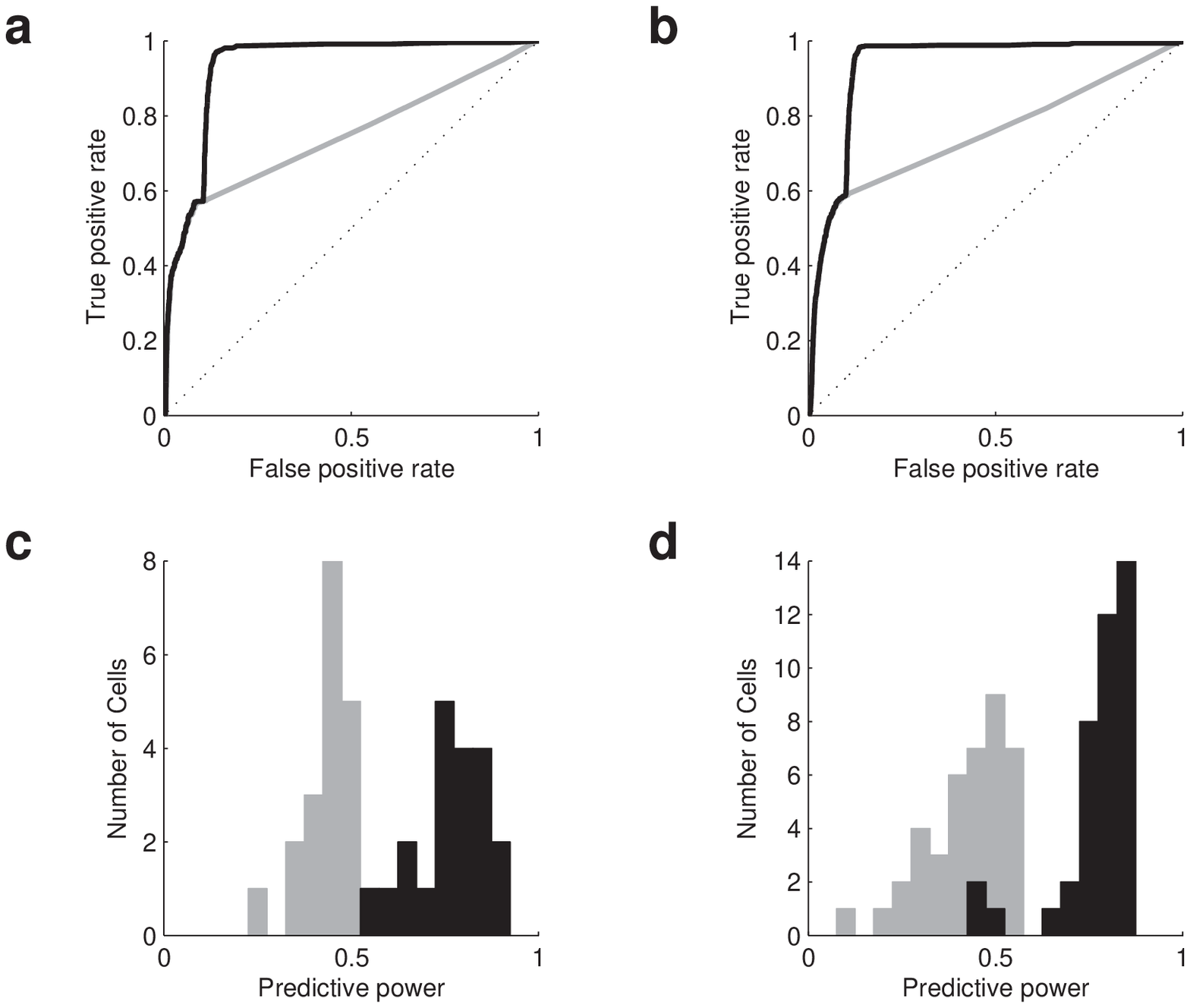}
\end{center}
\caption{Connection pattern 2: (a) ROC curves for our prediction scheme for a single cell (cell 12) from the 20 cell subset. (b) ROC curves for a single cell (cell 31) from the 40 cell subset. The gray trace in both cases results from thresholding over conditional probabilities generated by considering only the intrinsic spiking dynamics of the target cell. The black trace shows the case where the external cells are included in the analysis. (c,d) Histograms of the predictive powers obtained for the 20 cell subset (c), and the 40 cell subset (d). The gray bars indicate prediction based solely on the intrinsic spiking history of the target cell, whereas the black bars indicate predictive powers obtained by including the remainder of the ensemble.}
\label{Fig:Figure_9_ROC_PPHist}
\end{figure}

Applying thresholds to the conditional probabilities calculated as in Figures~\ref{Fig:Figure_6_condprobs_178} and \ref{Fig:Figure_7_condprobs_112}, we computed cross validated receiver operator characteristics (ROC) curves \citep{fawcett_06, truccolo+al_10}. ROC curves for two cells are shown in Figures~\ref{Fig:Figure_8_ROC_PPHist}(a,b) and \ref{Fig:Figure_9_ROC_PPHist}(a,b), taken from both the 20 and 40 cell subsets, for both connection patterns.

As in \cite{truccolo+al_10}, we measure predictive performance by computing the area under the ROC curve (AUC), as well as the chance level AUC (denoted by AUC*), obtained here by randomly shuffling the spike trains of the cells under consideration, and then define predictive power as $2(\textrm{AUC}-\textrm{AUC*})$. A predictive power of 0 signals chance level performance, whereas a predictive power close to 1 signals almost perfect performance. Histograms for the 20 cell case for both connection patterns are shown in Figures~\ref{Fig:Figure_8_ROC_PPHist}(c) and \ref{Fig:Figure_9_ROC_PPHist}(c) and for the 40 cell case in Figures~\ref{Fig:Figure_8_ROC_PPHist}(d) and \ref{Fig:Figure_9_ROC_PPHist}(d). Typical AUC* values were $\sim$ 0.50. We distinguish between the case where the optimization is done solely with the intrinsic spiking history of the target cell (shown in gray in Figures~\ref{Fig:Figure_8_ROC_PPHist} and \ref{Fig:Figure_9_ROC_PPHist}) and the case where the optimization is done with the inclusion of the remaining cells in the ensemble (shown in black in Figures~\ref{Fig:Figure_8_ROC_PPHist} and \ref{Fig:Figure_9_ROC_PPHist}).

For the 20 cell subset of connection pattern 1, considering only intrinsic spiking histories, we obtained predictive powers in the range of 0.32--0.45 with a mean of 0.38; in the full ensemble history case, we obtained values in the range of 0.47--0.64 with a mean of 0.57. For the 40 cell subset of connection pattern 1, predictive powers for the intrinsic spiking history case yielded a range of 0.34--0.48 with a mean of 0.41; whereas in the full ensemble history case, the predictive power ranged between 0.48--0.74 with a mean of 0.64. 

For the 20 cell subset of connection pattern 2, considering only intrinsic spiking histories, we obtained predictive powers in the range of 0.26--0.56 with a mean of 0.44; in the full ensemble history case, we obtained values in the range of 0.54--0.88 with a mean of 0.77. For the 40 cell subset of connection pattern 2, predictive powers for the intrinsic spiking history case yielded a range of 0.08--0.57 with a mean of 0.42; whereas in the full ensemble history case, the predictive power ranged between 0.46--0.87 with a mean of 0.78. We find that the pattern of connectivity chosen does impact the predictive power.

\begin{figure}
\begin{center}
\includegraphics[width=4in]{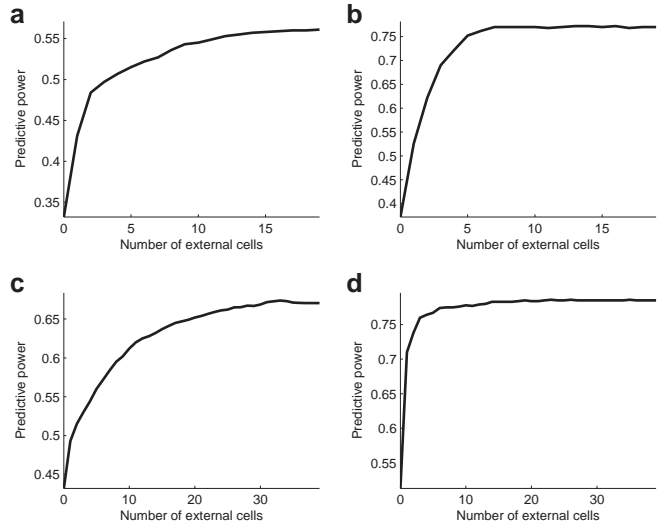}
\end{center}
\caption{Increment in the predictive power as a result of the addition of single cells, without replacement. (a) Cell 13 from connection pattern 1. (b) Cell 13 from connection pattern 2. (c) Cell 24 from connection pattern 1. (d) Cell 24 from connection pattern 2.}
\label{Fig:Figure_10_RawPPIncrement}
\end{figure}

A natural question which follows is whether one needs the full ensemble of cells to attain the full ensemble predictive power for any particular target cell under consideration. To address this question, we analyzed randomly chosen subsets of 5 cells from the 20 and 40 cell subsets for both connection patterns. For each cell analyzed, we monitored the change in predictive power which occurred as a result of adding a single cell from the remainder of the subset, without replacement, until all extrinsic cells from the subset had been added to the prediction scheme. The particular order in which cells were added was determined by an estimate of the potential influence they would have on the predictive power of the target cell. The results for one cell out of the chosen subset of 5, for each subset and from each connection pattern is shown in Figure~\ref{Fig:Figure_10_RawPPIncrement}. We note that a subset of the full ensemble of cells is sufficient to attain a sizable fraction of the maximal predictive power achieved with all cells from the ensemble included. This was borne out for all the cells analyzed.

\section{Discussion}

The existence of coordinated responses in small ensembles of nearby neurons has received renewed attention, in large part due to a growing number of recent experimental recordings employing a variety of theoretical formalisms \citep{schneidman+al_06, shlens+al_06, tang+al_08, yu+al_08, truccolo+al_10}. We investigated one aspect of this coordination through analyzing the predictive power inherent in the spiking histories of small ensembles of neurons in a computational model of cortex utilizing a point process framework \citep{truccolo+al_05, truccolo+al_10}. 

Our main finding is that surprising predictive power exists in small ensembles of modeled cortical neurons ($\sim 20 - 40$ neurons) where these cells are driven solely by random background inputs. A small fraction of the neurons in the model were subject to a synaptic current source large enough to drive an action potential in the neuron with high probability, with the times of synaptic current injection following a Poisson process. This background activity was chosen to mimic random inputs to our section of modeled cortex from external sources. This predictive power points to a strong redundancy in the neural code, and in particular highlights the possibility of error correcting properties at the neuronal level \citep{schneidman+al_06, truccolo+al_10}. 

We observe interesting effects which manifest due to the connectivity patterns studied. Connectivity pattern 1 is based on a more densely connected network, where layer II/III pyramidal cells generally synapse onto a greater number of layer II/III pyramidal cells in the network (at most 178 for pattern 1 vs 112 for pattern 2). The full ensemble predictive powers attained on average are however less for connectivity pattern 1 than those attained for pattern 2. A partial explanation for this fact is that for the less densely connected network (pattern 2), the pairwise correlation structure is stronger (comparing Figures~\ref{Fig:Figure_2_TrainsCorr_178} and~\ref{Fig:Figure_3_TrainsCorr_112}). In a similar vein, it seems that in the less connected, but more strongly correlated network (pattern 2), the addition of fewer extrinsic cells were needed to reach sizable fractions of the full ensemble predictive power (see Figure~\ref{Fig:Figure_10_RawPPIncrement}). Although the order in which cells were added wasn't necessarily the one which maximized the increase in predictive power at each addition, the fact of requiring only a subset of the full ensemble of cells was evident in all cases considered. The small oscillations in the predictive power one observes in Figure~\ref{Fig:Figure_10_RawPPIncrement} once one reaches close to the full ensemble predictive power (in Figure~\ref{Fig:Figure_10_RawPPIncrement}(b,c,d)) lie within our estimates for errors in computation of the predictive power ($\sigma\sim < 5\%$) \citep{hanley+mcneil_82}. It should be noted that the result that one can reach the full potential predictive power of an ensemble of neurons by observing subgroups of the ensemble has also been alluded to previously \citep{schneidman+al_06, tkacik+al_10}.

Our results add to the discussion first pointed out by \cite{truccolo+al_10}. There, it was shown that small ensembles of neurons in arm-related ares of sensorimotor cortex could reliably predict single-neuron spiking in humans and monkeys performing a range of sensorimotor tasks. They hypothesized that this predictive capability reflected behavior-related encoding, which could not be entirely accounted for by common inputs driving their networks. We show in the light of our above results that predictive capabilities lie in locally connected computational networks responding to sparsely distributed random background inputs.

Important differences between our computational model and the recordings outlined in \cite{truccolo+al_10} make a direct comparison difficult. We have chosen our ensemble of cells from the same layer of modeled cortex whereas the results of \cite{truccolo+al_10} most likely sample cells from multiple layers. The cross-sectional area from which we have sampled neurons is also likely to have been smaller. From the point of view of network activity, our model displays distinct periods of bursting, which together with our strategy for choosing our ensembles of neurons, might be responsible for the relatively stronger pairwise correlations we observe. Addressing some of these differences, such as randomizing over layers and sampling neurons spaced further apart (though of course our model is of limited size), should decrease the magnitude of the substantial predictive power we find, though we doubt this would eliminate its existence altogether. The question of the interpretation of our results is then intimately connected to the (open) question of how to interpret computational models themselves.

That we find predictive power in a separate system not ostensibly driven by strongly correlated inputs is clear. If one accepts our network and its inputs are not reflective of a piece of motor cortex (for example) performing behavior related computations, then we have at the very least constructed a realistic system in which significant predictive power exists. Predictive power is then not a unique signature of behavior though it of course may be implied by it. Furthermore, given the architecture of the computational model studied, our results hint that this predictive power may be a generic property of locally connected networks, and our model then provides an explicit, realistic structure within which one can begin to unravel the origins of this strongly correlated behavior. An alternative to this view is that our network is reflective of a piece of motor cortex performing behavior related computations, in which case our results provide further support to those of \cite{truccolo+al_10}, together then with an architecture through which such coordinated activity results.

In this way, we promote the interpretation that locally interconnected networks of neurons contain predictive power such that each unit in the network need not be monitored to obtain knowledge about the collective state of the ensemble. A number of avenues of pursuit present themselves in consequence. As pointed out in \cite{truccolo+al_10}, the distribution we have constructed does not constitute a {\em minimal} model for the system, it simply illustrates that predictive power exists. An interesting question is whether one can do better in terms of prediction through instituting a simpler model, whether it be constructed within the point process framework of \cite{truccolo+al_05, truccolo+al_10}, in terms of the maximum entropy approaches presented in the style of \cite{schneidman+al_06}, \cite{shlens+al_06}, \cite{tang+al_08}, \cite*{marre+al_09}, or indeed through lower dimensional state-space methods as discussed in \cite{stevenson+kording_11}. The point process formalism utilized for prediction was not explicitly tuned to necessarily generate optimal results, and the question of how well such inferential techniques do in terms of extracting available information from spike trains remains an interesting (open) question. In addition, the computational model chosen for analysis was also not explicitly tuned to generate the levels of predictive power we observed. How the parameter space of the computational model relates to the observed predictive power is also work for future consideration. The relationship of predictive power to connectivity is another interesting line of inquiry. One imagines there is a relationship between the temporal filters one attains and the connectivity of the network. Whether one can reverse the process and infer connectivity from predictive power would be of significant interest \citep{okatan+al_05, nykamp_07, kim+al_11, chen+al_11}. Indeed the setting of a reduced computational model, where connectivity and network architecture are under direct control, would be a promising place to begin.

\subsection*{Acknowledgments}

We thank Piotr Franaszczuk and Sridevi Sarma for discussions. Work at Harvard and at Johns Hopkins was supported in part by the Charles H. Hood Foundation and by the NIH (1K08NS066099-01A1).

\subsection*{Appendix: The computational model}

The computational model utilized here constitutes a version of that described in \cite{anderson+al_07, anderson+al_09} and \cite*{anderson+al_12}. Appendix A of \cite{anderson+al_07} contains a more complete description of a reduced version of the model under consideration, though for completeness we highlight portions of their discussion relevant to the model used here. This earlier version of the model contains the same cell types (as discussed below) and cell classes (layer II/III pyramidal, etc.), arranged in the same columnar structure as for the later version of the model (as featured in \cite{anderson+al_09, anderson+al_12}, and studied in this manuscript), together with similar intrinsic and extrinsic connectivity schemes. The main difference is the increased number of minicolumns arranged contiguously, taking the total number of cells from 16,384 to 65,536. We note also that the two models referred to above have been posted to the Yale SenseLab ModelDB database of computational neuroscience models (accession numbers 98902 and 141507) at http://senselab.med.yale.edu/modeldb/.

Each of the 65,536 neurons is modeled in single-compartment format in accord with the excitable membrane model of \cite{hodgkin+huxley_52}, as modified by \cite{avron+al_93} and \cite{avron_94}. The membrane potential $V$ for each cell varies as
\begin{equation}\label{MembPot}
C\frac{dV(t)}{dt} = I_{syn}(t) - I_{Na}(t)- I_{Ca}(t)- I_{K}(t)- I_{K(Ca)}(t)- I_{A}(t)- I_{L}(t),
\end{equation}
where $C$ is the membrane capacitance, $I_{syn}(t)$ is the input synaptic current, $I_{Na}(t)$ is the inward sodium current, $I_{Ca}(t)$ is the inward calcium current, $I_{K}(t)$ is the outward potassium current which includes a delayed rectifier current, $I_{K(Ca)}(t)$ is a calcium dependent potassium current, $I_{A}(t)$ is a transient potassium current, and $I_{L}(t)$ represents a leak current. The voltage dependence of these currents are presented in the appendix of \cite{anderson+al_07}, to which we refer the reader. The synaptic current $I_{syn}(t)$ for any neuron is determined through a weighted linear combination over all $N_{syn}$ synapses, where
\begin{equation}
I_{syn}(t) = \sum_{i = 1}^{N_{syn}}w_{i}g_{i}(t - \tau_{i})(V - E_{i}).
\end{equation}
The sum extends over all $N_{syn}$ synapses onto the neuron, $w_{i}$ is a weight which modulates the contribution of the $i$'th synapse, $g_{i}(t)$ is the conductance function for synapse $i$, which can depend upon a delay $\tau_{i}$. The term $E_{i}$ is the reversal potential utilized in the driving function for synapse $i$. The conductance functions $g_{i}(t)$ can be expressed as
\begin{equation}
g_{i}(t) = g_{syn}^{i}\sum_{j = 1}^{N(t)}\left(e^{\frac{t_{j}-t}{\tau_{d}}}-e^{\frac{t_{j}-t}{\tau_{o}}}\right),
\end{equation}
where $g_{syn}^{i}$ is a conductance constant which is adjusted to provide each postsynaptic potential with an equivalent amount of injected charge integrated over time. The sum runs over the non-negligible past action potential received by synapse $i$, and $\tau_{d}$ and $\tau_{o}$ are the decay and onset times for a given postsynaptic potential respectively.

Each neuron was modeled as one of three different types, regular spiking, intrinsic bursting, or fast spiking, depending upon its intrinsic electrophysiological properties. Model parameters which characterize each one of these types are presented in Table 2 of \cite{anderson+al_07}. The layer II/III pyramidal cell ensembles utilized in this paper were of the regular spiking type. Figure 1a of \cite{anderson+al_07} displays their response to an internal current injection, which notably exhibits a lack of self-excitation after the source is shut off. Specific parameter values governing their dynamics are presented in the second column of Table 2 in \cite{anderson+al_07}.

Connectivity between cells is described in detail in sections 2.1 and 2.2 of \cite{anderson+al_07}, and in \cite{anderson+al_09}. Most pertinent to our discussion here is the number of layer II/III pyramidal cells contacted by each layer II/III pyramidal cell in our chosen connectivity scheme. By way of example, for `connection pattern 1', layer II/III pyramidal cells make isotropic connections outside their minicolumn to a radius of 300$\mu\textrm{m}$, contacting a total of at most 178 layer II/III pyramidal cells along the way. The layer II/III pyramidal cells also contact (extrinsic to their own minicolumn) the layer IV stellate cells, layer V pyramidal cells, basket and chandelier cells. \cite{anderson+al_07} describe (in section 2) a basic connectivity pattern which is then varied via multiplicative factors acting on the total numbers of connections in the model to reproduce a variety of bursting behaviors with a range of frequencies. In this paper, we have set the relative numbers of inhibitory to excitatory connections to most resemble pattern C (Table 1, \cite{anderson+al_07}). For `connection pattern 2', the total number of contacted layer II/III pyramidal cells out to $300\mu\textrm{m}$ for each layer II/III pyramidal cell was reduced to at most 112. In both patterns, connections were reduced for cells within $300\mu\textrm{m}$ of the edge of the model to help mitigate boundary reflection effects.

\end{document}